\documentstyle[espart12]{article}
\def\p2t{p_{2\bot}}
\def\q2t{q_{2\bot}}
\begin{document}
\def\emline#1#2#3#4#5#6{%
       \put(#1,#2){\special{em:moveto}}%
       \put(#4,#5){\special{em:lineto}}}
\begin{center}
  \begin{Large}
  \begin{bf}
Small-x physics and
heavy quark photoproduction in the semihard approach
at HERA.
  \end{bf}

  \end{Large}
  \vspace{5mm}
  \begin{large}
    V.A.Saleev\\
  \end{large}
Samara State University, Samara 443011, Russia\\
  \vspace{5mm}
  \begin{large}
    N.P.Zotov\\
  \end{large}
Nuclear Physics Institute, Moscow State University,
      Moscow 119899, Russia\\

\end{center}
  \vspace{5mm}
\begin{abstract}
Processes of hevy quark photoproduction at HERA energies and beyond
are investigated using the semihard
($k_{\bot}$ factorization) approach. The virtuality and
longitudinal polarization of gluons in the photon - gluon subprocess
as well as the saturation effects in the gluon distribution
function at small $x$ have been taken into account.
The total cross sections, rapidity and $p_{\bot}$ distributions of
the charm and beauty quark photoproduction have been calculated.
We obtained the some  differences between the predictions
of the standard parton model and the semihard
approach used here.
\end{abstract}

\section{Introduction}
\par
 The heavy quark production at HERA is very intresting and important
 subject of study \cite{r1,r2,r3,r4}. Because it is
dominated by photon-gluon fusion subprocess (Fig.1) one can
study the gluon distribution functions $G(x, Q^2)$ in small $x$
region (roughly at $x > 10^{-4}$). Secondly this last issue
is important for physics at future colliders (such as LHC): many
processes at these colliders will be determined by small-x gluon
distributions.
\vspace{5mm}

\unitlength=1.00mm
\special{em:linewidth 1pt}
\linethickness{1pt}
\begin{picture}(104.00,16.00)
\emline{22.00}{1.00}{1}{37.00}{1.00}{2}
\emline{22.00}{1.00}{3}{22.00}{16.00}{4}
\emline{22.00}{16.00}{5}{37.00}{16.00}{6}
\emline{7.00}{16.00}{7}{9.00}{16.00}{8}
\emline{10.00}{16.00}{9}{12.00}{16.00}{10}
\emline{13.00}{16.00}{11}{15.00}{16.00}{12}
\emline{16.00}{16.00}{13}{18.00}{16.00}{14}
\emline{19.00}{16.00}{15}{21.00}{16.00}{16}
\emline{22.00}{1.00}{17}{20.00}{3.00}{18}
\emline{20.00}{3.00}{19}{20.00}{3.00}{20}
\emline{20.00}{3.00}{21}{18.00}{1.00}{22}
\emline{18.00}{1.00}{23}{16.00}{3.00}{24}
\emline{16.00}{3.00}{25}{14.00}{1.00}{26}
\emline{14.00}{1.00}{27}{12.00}{3.00}{28}
\emline{12.00}{3.00}{29}{10.00}{1.00}{30}
\emline{10.00}{1.00}{31}{8.00}{3.00}{32}
\emline{8.00}{3.00}{33}{6.00}{1.00}{34}
\put(1.00,16.00){\makebox(0,0)[cc]{$\gamma$}}
\put(1.00,1.00){\makebox(0,0)[cc]{g}}
\put(44.00,16.00){\makebox(0,0)[cc]{q}}
\put(44.00,1.00){\makebox(0,0)[cc]{$\bar q$}}
\emline{81.00}{1.00}{35}{81.00}{16.00}{36}
\emline{66.00}{16.00}{37}{68.00}{16.00}{38}
\emline{69.00}{16.00}{39}{71.00}{16.00}{40}
\emline{72.00}{16.00}{41}{74.00}{16.00}{42}
\emline{75.00}{16.00}{43}{77.00}{16.00}{44}
\emline{78.00}{16.00}{45}{80.00}{16.00}{46}
\emline{81.00}{1.00}{47}{79.00}{3.00}{48}
\emline{79.00}{3.00}{49}{79.00}{3.00}{50}
\emline{79.00}{3.00}{51}{77.00}{1.00}{52}
\emline{77.00}{1.00}{53}{75.00}{3.00}{54}
\emline{75.00}{3.00}{55}{73.00}{1.00}{56}
\emline{73.00}{1.00}{57}{71.00}{3.00}{58}
\emline{71.00}{3.00}{59}{69.00}{1.00}{60}
\emline{69.00}{1.00}{61}{67.00}{3.00}{62}
\emline{67.00}{3.00}{63}{65.00}{1.00}{64}
\put(60.00,16.00){\makebox(0,0)[cc]{$\gamma$}}
\put(60.00,1.00){\makebox(0,0)[cc]{g}}
\emline{81.00}{1.00}{65}{98.00}{16.00}{66}
\emline{81.00}{16.00}{67}{98.00}{0.00}{68}
\put(104.00,1.00){\makebox(0,0)[cc]{$\bar q$}}
\put(104.00,16.00){\makebox(0,0)[cc]{q}}
\end{picture}
\begin{center}
Fig.~1
\end{center}

 The heavy quark production processes are so-called semihard
processes ~\cite{r5} at HERA energies and beyond. By definition
in these processes a hard scattering scale $Q$ (or heavy quark
mass $M$) is large as compared to the $\Lambda_{QCD}$ parameter,
but $Q$ is much less than the total center of mass energies:
$\Lambda_{QCD} \ll Q \ll \sqrt s$.
Last condition implies that the processes occur in small $x$
region: $x\simeq M^2/s\ll 1$. In such case parturbative
QCD expansion have large cofficients $O(\ln^{n}(s/M^2))
\sim O(\ln
^{n}(1/x))$ besides the usual renormalization group ones
which are $O(\ln^{n}(Q^2/\Lambda^2))$.
So in perturbative QCD the heavy quark photoproduction cross section
as result of photon-gluon fusion processe has the form ~\cite{r6}:
$\sigma_{\gamma g} = \sigma^{o}_{\gamma g} + \alpha_{s}\sigma^{1}_
{\gamma g}$ + ..., where $\sigma^{1}_{\gamma g}$ was calculated
by Ellis and Nason ~\cite{r7}. The photon - gluon fusion cross
section in low order decreases vs $s$ at $s\rightarrow\infty:
\sigma^{o}_{\gamma g}\sim M^2/s\ln(s/Q^2)$.
But in the same limit $\sigma^{1}_{\gamma g}\rightarrow Const$,
because of that the heavy quark photoproduction cross section
is dominated by the contribution of the gluon exchange in the
$t$ - channel. That results  breakdown the standard
perturbative QCD expansion and the problem of summing up all
contributions of the order $(\alpha_{s} \ln(Q^2/\Lambda
^2))^{n},(\alpha_{s}\ln(1/x))^{n}$ and $(\alpha_{s}
\ln(Q^2/\Lambda^2)\ln(1/x)^{n}$ in
perturbative QCD appears in calculation of $\sigma_{\gamma g}$.
\par
It is known that summing up the terms of the order $(\alpha_{s}
\ln(Q^2/M^2))^{n}$ in leading logarithm approximation
(LLA) of perturbative QCD leads to the linear DGLAP evolution
equation for deep inelastic structure function ~\cite{r8}.
Resummation of the large contributions of the order of
$(\alpha_{s}\ln(1/x))^{n}$ leads to the BFKL evolution
equation ~\cite{r9} and its solution gives the drastical
increase of the gluon distribution: $xG(x,Q^2)\sim x^{-\omega_{0}},
\omega_{0}=(4N_c\ln2/\pi)\alpha_{s}(Q^2_{0})$.
The speed of the growth of the parton density as $x\rightarrow
0$ makes the parton-parton interactions very important,
which in turn makes the QCD so-called GLR evolution equation
 essentially nonlinear  ~\cite{r5}.

The growth of the parton density at $x\rightarrow 0$ and
interaction between partons induces substantial screening
(shadowing) corrections which restore the unitarity constrains
for deep inelastic structure functions (in particular for
a gluon distribution) in small $x$ region ~\cite{r10}.
These facts break the assumption of the standard parton model (SPM)
about $x$ and transverse momentum factorization for a parton
distribution functions. We should deal with the transverse
momentum factorization ($k_{\bot}$ factorization) theory
{}~\cite{r11,r12}. Making allowance for screening corrections at
small $x$ we have
 so-called semihard approach \cite{r5,r10}, which take into account
the virtuality of gluons, their transverse momenta as well as
the longitudional polarizations of initial gluons.

In semihard approach ~\cite{r5,r10} screening corrections stop
the growth of the gluon distribution at $x\rightarrow 0$.
This effect was interpreted as saturation of the parton
(gluon) density: gluon distribution function $xG(x,Q^2)$ becames
proportional to $Q^2R^2$ at $Q^2\leq q^2_{0}(x)$ and $\sigma
\sim \frac{1}{Q^2}xG(x,Q^2)\sim R^2$. The parameters $R$ and
$q^2_{0}(x)$ can be considered as new phenomenological
parameters: the $R$ is related to size of hadron or of black
spots in a hadron and the parameter $q^2_{0}$ is a typical
transverse momentum of partons in the parton cascade of
a hadron, which leads to natural infrared cut-off in semihard
processes. The parameter $q^2_{0}$ increases with growth of $s$
{}~\cite{r10,r13}. So, because of the high parton density in small $x$
region a standard methods of perturbtive QCD can not be used
{}~\cite{r5,r10,r13,r14}. In this point the semihard approach
{}~\cite{r5,r10} differs from the works ~\cite{r11,r12,r15} where
after choice of structure function at starting value $Q^2_{0}$ the
QCD evolution are calculated using certain equations.

In the region, where the transverse mass of heavy quark $M_{\bot}
<<\sqrt{s}$, one need take into account the dependence
of the photon-gluon fusion cross section on the virtuality and
polarizations of initial gluon. Thus in semihard approach the
matrix elements of this subprocess differs from ones of SPM
(see for example ~\cite{r6,r12}).

We would like to remark that $\alpha_s$ corrections in the semihard
approach look quite different in comparison with usual \cite{r6,r7}
and $k_{\bot}$ factorization \cite{r11,r12} ones. This corrections
have been taken into account using new gluon density $\Phi(x,\vec
q_{\bot}^2)$ (see below) which was calculated by Gribov, Levin and
Ryskin \cite{r5} in LLA where all large logs (of the types $ln(1/x)$
and $ln(Q^2/\lambda^2)$) were summed up. The another corrections which
give contributions to $K-$factor have been taken into account in the
normalization of function $xG(x,Q^2)$ (see below also) \cite{r10}.

As far as the SPM calculations of the next-to-leading (NLO)
cross section
to the photoproduction of heavy quarks that the review of its
may found in the papers ~\cite{r2,r16}. The results ~\cite{r7} have been
confirmed by Smith and Van Neerven ~\cite{r17}. The futher results
for the electro- and photoproduction of heavy quarks was obtained in
Refs. ~\cite{r18,r19}. Authors in ~\cite{r16}
notes that for beauty quark production the NLO corrections
are large and various estimates of corrections lead to theoretical
uncertainties of the order of a factor of 2 to 3.
As was estimated
in ~\cite{r21} the total cross section for beauty quark production
at HERA will be only a few tens of percent large than the one-loop
results ~\cite{r7}.

For charm quark a theoretical uncertainties are even the higher.
It is known that there is also the strong mass dependence of
the results of calculations for charm quark production.
  Since the mass of the charm quark is small for perturbative
QCD calculations the resumme procedure ~\cite{r11} is not
applicable for charm quark production ~\cite{r21}.

In Refs.~\cite{r22,r23} we used the semihard approach in order to
calculate the total and differential cross sections of the heavy
quarkonium, $J/\Psi$ and $\Upsilon$, photoproduction. We obtained
the remarkable difference between the predictions of the semihard
approach and the SPM especially for $p_{\bot}$- and $z$-distributions
of the $J/\Psi$ mesons at HERA.

In present paper we investigate the open heavy quark production
processes in the semihard approach, which was used early in
Ref.~\cite{r10} for calculation of heavy quark production rates
at hadron colliders and for prediction of $J/\Psi$ and$\Upsilon$
photoproduction cross sections at high energies in Refs.~\cite{r22,r23}.

\section{QCD Cross Section for Heavy Quark Electroproduction}
\par
 We calculate the total and differential cross sections (the
$p_{\bot}$ and rapidity distributions) of charm and beauty quark
photoproduction
via the photon-gluon fusion QCD subprocess (Fig.1) in the framework
of the semihard ($k_{\bot}$ factorization) approach ~\cite{r4,r10}.
First of all we take into account the transverse momentum of gluon
$\vec{q}_{2\bot}$, its  virtuality $q^2_{2}=-\vec{q}^2_{2\bot}$ and
the alignment of the polarization vectors along the transverse momentum
such as $\epsilon_{\mu} = q_{2\bot\mu}/\mid\vec{q}_{2\bot}\mid$
{}~\cite{r10,r12}.
\vspace{5mm}

\unitlength=1.00mm
\special{em:linewidth 1pt}
\linethickness{1pt}
\begin{picture}(86.00,40.00)
\emline{41.00}{35.00}{1}{56.00}{35.00}{2}
\emline{56.00}{35.00}{3}{71.00}{40.00}{4}
\emline{41.00}{5.00}{5}{56.00}{5.00}{6}
\put(58.00,5.00){\circle*{5.20}}
\put(64.00,20.00){\circle{8.00}}
\emline{56.00}{35.00}{7}{58.00}{32.00}{8}
\emline{59.00}{31.00}{9}{61.00}{28.00}{10}
\emline{62.00}{27.00}{11}{64.00}{24.00}{12}
\emline{64.00}{16.00}{13}{64.00}{13.00}{14}
\emline{64.00}{13.00}{15}{62.00}{13.00}{16}
\emline{62.00}{13.00}{17}{63.00}{10.00}{18}
\emline{63.00}{10.00}{19}{61.00}{10.00}{20}
\emline{61.00}{10.00}{21}{62.00}{7.00}{22}
\emline{62.00}{7.00}{23}{60.00}{7.00}{24}
\emline{67.00}{23.00}{25}{81.00}{25.00}{26}
\emline{67.00}{17.00}{27}{81.00}{14.00}{28}
\put(33.00,35.00){\makebox(0,0)[cc]{$e(P_1)$}}
\put(33.00,4.00){\makebox(0,0)[cc]{$p(P_2)$}}
\put(54.00,29.00){\makebox(0,0)[cc]{$q_1$}}
\put(54.00,12.00){\makebox(0,0)[cc]{$q_2$}}
\put(86.00,10.00){\makebox(0,0)[cc]{$\bar q(p_2)$}}
\put(86.00,27.00){\makebox(0,0)[cc]{$q(p_1)$}}
\emline{61.00}{5.00}{29}{76.00}{5.00}{30}
\emline{60.00}{3.00}{31}{76.00}{3.00}{32}
\end{picture}

\begin{center}
Fig.~2
\end{center}

Let us define Sudakov variables of the process $ep\to Q\bar Q X$
(Fig.2):
\begin{eqnarray}
p_1&=&\alpha_1 P_1+\beta_1 P_2+p_{1\bot},\qquad
p_2=\alpha_2 P_1+\beta_2 P_2+\p2t,\nonumber\\
q_1&=&x_1P_1+q_{1\bot},\qquad
q_2=x_2P_2+\q2t,
\end{eqnarray}
where
$$p_1^2=p_2^2=M^2,\qquad q_1^2=q_{1\bot}^2,\qquad q_2^2=\q2t^2,$$
$p_1$ and $p_2$ are 4-momenta of the heavy quarks, $q_1$ is 4-momentum
 of the photon, $q_2$ is 4-momentum of the gluon, $p_{1\bot},~~p_{2\bot},~~
 q_{1\bot},~~q_{2\bot}$ are transverse 4-momenta of these ones.
 In the center of mass frame of colliding particles we can write
 $P_1=(E,0,0,E)$, $P_2=(E,0,0,-E)$, where $E=\sqrt s/2$, $P_1^2=P_2^2=0$
  and $(P_1P_2)=s/2$.
Sudakov variables are expressed  as follows:
\begin{eqnarray}
\alpha_1 &=&\frac{M_{1\bot}}{\sqrt s}\exp(y_1^\ast),\qquad
\alpha_2 =\frac{M_{2\bot}}{\sqrt s}\exp(y_2^\ast),\nonumber\\
\beta_1 &=&\frac{M_{1\bot}}{\sqrt s}\exp(-y_1^\ast),\qquad
\beta_2 =\frac{M_{2\bot}}{\sqrt s}\exp(-y_2^\ast,),
\end{eqnarray}
where $M_{1,2\bot}^2=M^2+p_{1,2\bot}^2$, $y_{1,2}^{\ast}$ are rapidities
of heavy quarks, M is heavy quark mass.

 From conservation laws we can easy obtain following conditions:
\begin{equation}
q_{1\bot}+\q2t=p_{1\bot}+\p2t,\qquad
 x_1=\alpha_1 +\alpha_2,\qquad
 x_2=\beta_1 +\beta_2.
\end{equation}
The differential cross section of heavy quarks electroproduction has form:
\begin{equation}
\frac{d\sigma}{d^2p_{1\bot}}(ep\to Q\bar Q X)=
\int dy_1^{\ast}dy_2^{\ast}\frac{d^2q_{1\bot}}{\pi}\frac{d^2q_{2\bot}}{\pi}
\frac{|\bar M|^2\Phi_e(x_1,q_{1\bot}^2)\Phi_p(x_2,q_{2\bot}^2)}
 {16\pi^2(x_1x_2s)^2}
 \end{equation}
For photoproduction process it reads:
\begin{equation}
\frac{d\sigma}{d^2p_{1\bot}}(\gamma p\to Q\bar Q X)=
\int dy_1^{\ast}\frac{d^2q_{2\bot}}{\pi}
\frac{\Phi_p(x_2,q_{2\bot}^2)|\bar M|^2}
 {16\pi^2(sx_2)^2\alpha_2}
\end{equation}
We use generalized gluon structure function of a proton
 $\Phi_p(x_2,q_{2\bot}^2)$
which is obtained in semihard approach.  When
integrated over transverse momentum $\vec q_{2\bot}$
$(q_{2\bot}=(0,\vec q_{2\bot},0))$ of gluon up to some limit $Q^2$ it
becomes the  usual structure
function giving the gluon momentum fraction distribution at  scale
$Q^2$:
\begin {equation}
\int\limits^{Q^2}\Phi_p
(x,q_{2\bot}^2)d\vec q_{2\bot}^2=xG_p(x,Q^2).
\end{equation}
We use in our calculation following phenomenological parameterization
{}~\cite{r10}:
\begin{equation}
\Phi_p(x,q_{\bot}^2)=C\frac{0.05}{x+0.05}(1-x)^3f_p(x,q_{\bot}^2),
\end{equation}
where
\begin{eqnarray}
f_p&=&1,\qquad q_{\bot}^2\le q_0^2(x)\nonumber\\
f_p&=&(\frac{q_0^2(x)}{\vec q_{\bot}^2})^2, \qquad \vec q_{\bot}^2>q_0^2(x),
\end{eqnarray}
and $q^2_{0}(x) = Q^2_{0} + \Lambda^2\exp( 3.56 \sqrt{ \ln(x_0/x)})$,
 $Q_{0}^2 = 2 \mbox{GeV}^2$, $\Lambda = 56$ MeV, $x_{0}$ = 1/3.
The normalization factor $C\simeq 0.97$ mb of the structure function
$\Phi_p(x,\vec q_{2\bot}^2)$ was obtained in ~\cite{r10} where
$b\bar b$-pair
production at Tevatron energy was described.

The $\Phi_e(x_1,q_{1\bot}^2)$ is well known virtual photon spectrum
in Weizsacker-Williams approximation ~\cite{r24} before the integration
over  $\vec q_{\bot}^2$:
\begin{equation}
 \Phi(x_1,\vec q_{\bot}^2)=\frac{\alpha}{2\pi}[\frac{1+(1-x_1)^2}
  {x_1\vec q_{\bot}^2}-\frac{2m_e^2x_1}{\vec q_{\bot}^4}].
\end{equation}

The effective gluon distribution $xG(x,Q^2)$ obtained from (7)
increases at not very small $x$ $(0.01 < x < 0.15)$ as $x^{-\omega_0}$,
where $\omega_0 = 0.5$ corresponds to the BFKL Pomeron singularity
{}~\cite{r9}.This increases continuously up to $x = x_0$, where
$x_0$ is a solution of the equation $q^2_0(x_0) = Q^2$. In the region
$x < x_0$ there is the saturation of the gluon distribution:
$xG(x,Q^2)\simeq CQ^2$.

The square of matrix element of partonic subprocess $\gamma^{\ast} g^{\ast}
\to Q\bar Q$
can be written as follows:
\begin{eqnarray}
|M|^2&=&16\pi^2e_Q^2\alpha_s\alpha(x_1x_2s)^2[
 \frac{1}{(\hat u-M^2)(\hat t-M^2)}-\nonumber\\
&&
\frac{1}{q_{1\bot}^2q_{2\bot}^2}
(1+\frac{\alpha_2\beta_1s}{\hat t-M^2}+
 \frac{\alpha_1\beta_2s}{\hat u-M^2})^2]
\end{eqnarray}
For real photon and off-shell gluon it reads:
\begin{eqnarray}
|M|^2&=&16\pi^2e_Q^2\alpha_s\alpha(x_2s)^2[\frac{\alpha_1^2+\alpha_2^2}
 {(\hat t-M^2)(\hat u-M^2)}+\nonumber\\
   && \frac{2M^2}{q_{2\bot}^2}
(\frac{\alpha_1}{\hat u-M^2}+
 \frac{\alpha_2}{\hat t-M^2})^2],
\end{eqnarray}
where $\alpha_2=1-\alpha_1$ and $\hat s,~\hat t,~~\hat u$ are usual
Mandelstam variables of partonic subprocess
\begin{eqnarray}
 \hat s&=&(p_1+p_2)^2=(q_1+q_2)^2,\qquad
 \hat t=(p_1-q_1)^2=(p_2-q_2)^2,\\
 \hat u&=&(p_1-q_2)^2=(p_2-q_1)^2,\qquad
 \hat s+\hat t+\hat u=2M^2+q_{1\bot}^2+q_{2\bot}^2. \nonumber
\end{eqnarray}

\section{Discussion of the Results}
\par
The results of our calculations for the total cross sections of
$c$- and $b$-quark photoproduction are shown in Fig.3
(solid curves). Dashed curves correspond to the SPM predictions with
the GRV parametrization of the gluon distribution~\cite{r25}.
The results of calculations for the SPM are shown without K-factor,
wich have the typical value K = 2 for hard scattering processes.
(In the semihard approach K-factor is absent ~\cite{r10,r23}).
The K-factor in the SPM may change the relation between the
results of calculations in the semihard approach and SPM.
But independently from it we see that the saturation effects
more clearly are pronounced for charm quark photoproduction
(at $\sqrt{s_{\gamma p}}\sim 500$ GeV). In any case (with or
without K-factor in SPM) the cross section for beauty quark
photoproduction in the semihard approach is the higher than
one in the SPM at $\sqrt{s_{\gamma p}} > 200$ GeV.

The $p_{\bot}$ distributions for $c$- and $b$-quark photoproduction
in the semihard approach (solid curves) and in the SPM
(dashed curves) are shown in Fig.4.
 The curves are obtained in the semihard
approach for charm quark photoproduction show the saturation effects
in low $p_{\bot}$ region ($p_{\bot} < 2$ Gev/c). In middle
$p_{\bot}$ region (2 GeV/c $< p_{\bot} < 15$ GeV/c) the
heavy quark photoproduction $p_{\bot}$ distributions in
the semihard approach are the higher ones of the SPM (with GRV
paramatrization of gluon distribution). At high $p_{\bot} >
15$ GeV/c we have contrary relation between $p_{\bot}$
distributions in the semihard approach and SPM.

Thus the SPM leads to over-estimatedcross section in the low
$p_{\bot}$ region and under-estimated one in the middle
$p_{\bot}$ region as was noted as early as in Refs.
{}~\cite{r10,r15} (see also ~\cite{r23}). This
behavior of $p_{\bot}$ distributions in the $k_{\bot}$
factorization approach is result from the off mass shell
subprocess cross section ~\cite{r15} as well as the saturation
effects of gluon structure function in semihard approach
{}~\cite{r10}.

Fig.5 show the comparison of the results for heavy quark rapidity
distribution (in the photon-proton center of mass frame)
in the different models:
solid curve shows the $y$ distribution in the
semihard approach,  dashed curve shows one in SPM. The discussed
above effects are sufficiently large near the kinematic boundaries,
i.e. at big value of $|y^{\ast}|$. We see that the difference between solid
and dushed curvers can't be degrade at all $y^{\ast}$ via change of
normalization of SPM prediction.

\section{Conclusions}
We shown that the semihard approach leads to the saturation effects
for the total cross section of charm quark photoproduction
at available energies and
predicts the remarkable difference for rapidity and transverse
momentum distributions of charm and beauty quark photoproduction,
which can be study already at HERA $ep$ collider.

\vspace{3mm}
{\bf Acknowledgements}
\vspace{3mm}

This  research was supported by the Russian
  Foundation of Basic Research (Grant 93-02-3545).
Authors would like to thank J.Bartels, S.Catani,
G.Ingelman, H.Jung,
J.Lim, M.G.Ryskin and A.P.Martynenko for fruitfull discussions
of the obtained results.
One of us (N.Z.) gratefully acknowladges W.Buchmuller,
G.Ingelman, R.Klanner, P.Zerwas and DESY directorate for
hospitality and support at DESY, where part of this work
was done.

                %=======================================================
{\bf Figure captions}
\begin{enumerate}
\item
QCD diagrams for open heavy quark photoproduction subprocesses
\item
Diagram for heavy quark elecrtoproduction
\item
The total cross section for open charm and beauty quark photoproduction:
solid curve - the semihard approach, dashed curve - the SPM
\item
The $p_{\bot}$ distribution for charm and beauty quark photoproduction
at $\sqrt{s_{\gamma p}} = 200$ GeV: curves as in Fig.3
\item
The $y^{*}$ distribution for charm and beauty quark photoproduction:
at $\sqrt{s_{\gamma p}} = 200$ GeV: curves as in Fig.3
\end{enumerate}
\end{document}